\documentstyle[aps,preprint,tighten]{revtex}

\begin{document}

\draft

\title{Comment on ``Quantum mechanics of an electron in a 
homogeneous magnetic field and a singular magnetic flux tube''}
\author{R. M. Cavalcanti\footnote{E-mail:rmoritz@fma.if.usp.br}}
\address{Instituto de F\'{\i}sica, Universidade de S\~ao Paulo,
Cx.\ Postal 66318, 05315-970, S\~ao Paulo, SP, Brazil}
\date{\today}
\maketitle

\begin{abstract}

Recently Thienel [{\it Ann.\ Phys.\ (N.Y.)} {\bf 280} (2000), 140]
investigated the Pauli equation for an
electron moving in a plane under the influence of a perpendicular
magnetic field which is the sum of a uniform field and a
singular flux tube. 
Here we criticise his claim that one cannot properly
solve this equation by treating the singular flux
tube as the limiting case of a flux tube of finite size.

\end{abstract}

\pacs{}

%%%%%%%%%%%%%%%%%%%%%%%%%%%%%%%%%%%%%%%%%%%%%%%%%%%%%%%%%%%%%%

The Pauli Hamiltonian for an electron (of mass $M$, charge $-|e|$
and $g$-factor 2) moving in the $(x,y)$-plane
under the influence of a magnetic field pointing in the 
$z$-direction is given by
\begin{equation}
\label{Pauli}
H=\frac{1}{2M}\left({\bf p}+\frac{|e|}{c}\,{\bf A}\right)^2
+\frac{|e|\hbar}{Mc}\,B_zS_z.
\end{equation}
Thienel \cite{Thienel} has recently investigated 
the eigenvalue problem for this Hamiltonian
in the case of a magnetic field which is the sum of
a uniform field and a singular flux tube,
\begin{equation}
\label{Bz}
B_z({\bf r})=B+\alpha\Phi\delta^2({\bf r})\qquad
\left(B>0;\,\Phi\equiv 2\pi\hbar c/|e|\right).
\end{equation}
He claims that standard approaches to this problem fail. 
The purpose of this Comment is to show not only that they
do work, they are also simpler than his
alternative method. 

As Thienel, we choose the vector potential in the symmetric gauge,
$$
{\bf A}(r)=\left(\frac{Br}{2}
+\frac{\alpha\Phi}{2\pi r}\right){\bf e}_{\varphi},
$$
and use magnetic units (where the unit of
length is $\lambda=(\Phi/\pi B)^{1/2}$ and the
unit of energy is $\hbar\omega$, with $\omega=|e|B/Mc$
the Larmor frequency). Then we can rewrite (\ref{Pauli}) as
\begin{equation}
H=-\frac{1}{4r}\,\frac{\partial}{\partial r}\left(r\,
\frac{\partial}{\partial r}\right)-\frac{1}{4r^2}
\left(\frac{\partial}{\partial\varphi}+i\alpha\right)^2
-\frac{i}{2}\left(\frac{\partial}{\partial\varphi}
+i\alpha\right)+\frac{1}{4}\,r^2+\left[1+\frac{\alpha}{2r}\,
\delta(r)\right]S_z
\end{equation}
(our $r$ corresponds to his $\tilde{r}$).
Since $H$, $L_z$ and $S_z$
commute with each other they can be diagonalized
simultaneously, so we can write the eigenfunctions of
$H$ as $\Psi_{E,m,\sigma}(r,\varphi)
=\psi_{E,m,\sigma}(r)\,e^{im\varphi}\,|\sigma\rangle$,
with $m$ an integer and  
$S_z\,|\sigma\rangle=\sigma\,|\sigma\rangle$, 
$\sigma=\pm 1/2$. Solving the resulting differential 
equation for $\psi_{E,m,\sigma}(r)$ and demanding that
$\psi_{E,m,\sigma}(r)\to 0$ as $r\to\infty$
we finally obtain
\begin{equation}
\label{Psi}
\Psi_{E,m,\sigma}(r,\varphi)={\cal N}\,r^{|m+\alpha|}\,e^{-r^2/2}\,
U(\xi,|m+\alpha|+1,r^2)\,e^{im\varphi}\,|\sigma\rangle,
\end{equation}
where $U(a,b,z)$ is one of Kummer's functions \cite{Abramowitz},
${\cal N}$ is a normalization constant and
\begin{equation}
\label{xi}
\xi\equiv\frac{1}{2}\,(|m+\alpha|+m+\alpha+1+2\sigma)-E.
\end{equation}

In order to determine the possible values of $E$ 
we need to know the correct boundary condition at the origin.
This problem was examined by Hagen \cite{Hagen} 
and G\'ornicki \cite{Gornicki}
in the case of a pure Aharonov-Bohm potential (i.e., with $B=0$). 
By treating the singular
flux tube as the limiting case of a flux tube of finite size,
they obtained the following result: the eigenfuntions
corresponding to the spin component which ``sees'' a repulsive 
delta-function
potential at the origin (i.e., $\sigma=+1/2$ if $\alpha>0$,
$\sigma=-1/2$ if $\alpha<0$) must be regular there.
(By contrast, Thienel requires, without justification, that 
they vanish at the origin. This is the reason why a
vacancy line occurs in his $(E,m+\sigma)$-plane --- see
FIG.\ 1 of Ref.\ \cite{Thienel} --- for integer $\alpha\ne 0$,
whereas in our solution no such vacancy line occurs.)
The supersymmetry of the Pauli Hamiltonian \cite{Thienel,Casher}
determines the eigenfunctions corresponding to the other spin 
component. 

A few remarks are in order here:
(i) although such a boundary condition was derived
for the the Dirac equation, one can easily show that 
it also holds for the Pauli equation 
(with $g=2$) \cite{Casher,Jackiw};
(ii) the presence of background smooth magnetic field does
not alter the boundary condition at the origin, as it does not add any
singular term to the Hamiltonian.

Let us first consider the case $\alpha>0$.
Then $\Psi_{E,m,1/2}$ must be regular at the origin,
which occurs only if $\xi=-n$ $(n=0,1,2\ldots)$,
for then
\begin{equation}
\label{Laguerre}
U(-n,|m+\alpha|+1,r^2)=(-1)^n\,n!\,L_n^{|m+\alpha|}(r^2),
\end{equation}
where $L_n^a(z)$ is the associated Laguerre polynomial 
\cite{Abramowitz}. Combining this with (\ref{Psi}) and
(\ref{xi}) and normalizing $\Psi_{E,m,1/2}$ to unity
we thus obtain
\begin{eqnarray}
& &\Psi_{n,m,1/2}(r,\varphi)
=\sqrt{\frac{\Gamma(n+1)}{\pi\,\Gamma(|m+\alpha|+n+1)}}\,
r^{|m+\alpha|}\,e^{-r^2/2}\,
L_n^{|m+\alpha|}(r^2)\,e^{im\varphi}\,|+\rangle,
\label{Psi+}
\\
& &E_{n,m,1/2}=n+1+\frac{1}{2}\,(|m+\alpha|+m+\alpha)
\qquad(n=0,1,2\ldots; m=0,\pm 1,\pm 2\ldots).
\end{eqnarray}
For each of these states there is a superpartner
with the same energy and opposite spin, obtained by
applying the supercharge $Q^{\dag}$ 
(Eq.\ (10) of Ref.\ \cite{Thienel}) 
to (\ref{Psi+}):
\begin{eqnarray}
\Psi_{n,m+1,-1/2}(r,\varphi)
&=&E_{n,m,1/2}^{-1/2}\,Q^{\dag}\,\Psi_{n,m,1/2}(r,\varphi)
\nonumber \\
&=&\frac{1}{2}\sqrt{\frac{\Gamma(n+1)}{\pi\,E_{n,m,1/2}\,
\Gamma(|m+\alpha|+n+1)}}
\nonumber \\
& &\times\left(-\frac{\partial}{\partial r}+\frac{m+\alpha}{r}
+r\right)r^{|m+\alpha|}\,e^{-r^2/2}\,
L_n^{|m+\alpha|}(r^2)\,e^{i(m+1)\varphi}\,|-\rangle.
\label{Psi-}
\end{eqnarray}
The factor $E_{n,m,1/2}^{-1/2}$ ensures proper normalization:
\begin{eqnarray*}
\langle\Psi_{n,m+1,-1/2}|\Psi_{n,m+1,-1/2}\rangle
&=&E_{n,m,1/2}^{-1}\,\langle\Psi_{n,m,1/2}|\,QQ^{\dag}\,
|\Psi_{n,m,1/2}\rangle 
\\
&=&E_{n,m,1/2}^{-1}\,\langle\Psi_{n,m,1/2}|\,(H-Q^{\dag}Q)\,
|\Psi_{n,m,1/2}\rangle=1.
\end{eqnarray*}
The eigenstates with zero energy are anihilated by both
supercharges, 
$$
Q\,|E=0\rangle=Q^{\dag}\,|E=0\rangle=0.
$$
They are given by
\begin{equation}
\label{Psi0}
\Psi_{E=0,m,-1/2}(r,\varphi)
=\frac{r^{-(m+\alpha)}\,e^{-r^2/2}\,e^{im\varphi}\,|-\rangle}
{\sqrt{\pi\,\Gamma(-m-\alpha+1)}};
\end{equation}
square integrability requires $m+\alpha<1$.

If $\alpha<0$ it is $\Psi_{E,m,-1/2}$ which must be regular
at $r=0$. Thus
\begin{eqnarray}
& &\Psi_{n,m,-1/2}(r,\varphi)
=\sqrt{\frac{\Gamma(n+1)}{\pi\,\Gamma(|m+\alpha|+n+1)}}\,
r^{|m+\alpha|}\,e^{-r^2/2}\,
L_n^{|m+\alpha|}(r^2)\,e^{im\varphi}\,|-\rangle,
\\
& &E_{n,m,-1/2}=n+\frac{1}{2}\,(|m+\alpha|+m+\alpha)
\qquad(n=0,1,2\ldots; m=0,\pm 1,\pm 2\ldots).
\end{eqnarray}
The zero modes are already included among these states 
$(n=0, m+\alpha\le 0)$. The spin-up states are obtained
by applying the supercharge $Q$ (Eq.\ (9) of Ref.\ \cite{Thienel})
to the spin-down states with nonzero energy (the factor 
$E_{n,m,-1/2}^{-1/2}$ ensures proper normalization):
\begin{eqnarray}
\Psi_{n,m-1,1/2}(r,\varphi)
&=&E_{n,m,-1/2}^{-1/2}\,Q\,\Psi_{n,m,-1/2}(r,\varphi)
\nonumber \\
&=&\frac{1}{2}\sqrt{\frac{\Gamma(n+1)}{\pi\,E_{n,m,-1/2}\,
\Gamma(|m+\alpha|+n+1)}}
\nonumber \\
& &\times\left(\frac{\partial}{\partial r}+\frac{m+\alpha}{r}
+r\right)r^{|m+\alpha|}\,e^{-r^2/2}\,
L_n^{|m+\alpha|}(r^2)\,e^{i(m-1)\varphi}\,|+\rangle.
\end{eqnarray}

The same results can be obtained by treating the singular
flux tube as the limiting case of a flux tube of finite
size in the presence of a background homogeneous magnetic
field. The calculations, however, are more complicated
and not illuminating. Here I shall only point out to
the origin of Thienel's wrong conclusion that such an
approach fails. First of all, we note that the correct form of 
$\Psi$ outside the flux tube is given by (\ref{Psi}).
Then, by demanding continuity of
$\partial\Psi/\partial r$ at the border of
the tube, one is led to what is essentially his Eq.\ (17) 
multiplied by
$R^{|m+\alpha|-1}\,e^{-R^2/2}\,U(\xi,|m+\alpha|+1,R^2)$.
As a function of $\xi$, $U$
has an infinite number of zeros, which were completely
overlooked by Thienel. [One can show, in particular,
that the zeros $\xi_n$ of $U$
satisfy $\lim_{R\to 0}\xi_n=-n$ ($n=0,1,2\ldots$).
This follows from the asymptotic behavior of $U$ for
small $R$ \cite{Abramowitz},
$$
U(\xi,|m+\alpha|+1,R^2)\stackrel{R\to 0}{\sim}
-\frac{\Gamma(|m+\alpha|)}{\Gamma(\xi)}\,R^{-2|m+\alpha|}
$$
(valid for $\xi\ne -n$ and $m+\alpha\ne 0$), the fact that
$\Gamma(-n+\epsilon)$ and $\Gamma(-n-\epsilon)$ have
opposite signs for $n=0,1,2\ldots$ and $0<\epsilon<1$,
and the continuity of $U$ as a function of $\xi$.]

%%%%%%%%%%%%%%%%%%%%%%%%%%%%%%%%%%%%%%%%%%%%%%%%%%%%%%%%%%%%%%

\acknowledgments

I thank Adilson Jos\'e da Silva and Marcelo Gomes
for a critical reading of this paper.
This work was supported by FAPESP.

%%%%%%%%%%%%%%%%%%%%%%%%%%%%%%%%%%%%%%%%%%%%%%%%%%%%%%%%%%%%%%

\end{document}